# Navigating the complex nexus: cybersecurity in political landscapes

Mike Nkongolo[1*]

[1*]Department of Informatics, University of Pretoria, Hatfield 0028, Pretoria, Gauteng, South-Africa.

E-mail: mike.wankongolo@up.ac.za

**Abstract**

Cybersecurity in politics has emerged as a critical and intricate realm intersecting technology, governance, and international relations. In today's interconnected digital context, political entities confront unparalleled challenges in securing sensitive data, upholding democratic procedures, and countering cyber threats. This study delves into the multifaceted landscape of political cybersecurity, examining the evolving landscape of cyberattacks, their impact on political stability, and strategies for bolstering digital resilience. The intricate interplay between state-sponsored hacking, disinformation campaigns, and eroding public trust underscores the imperative for robust cybersecurity measures to safeguard political system integrity. Through an extensive exploration of real-world case studies, policy frameworks, and collaborative initiatives, this research illuminates the intricate network of technological vulnerabilities, geopolitical dynamics, and ethical concerns that shape the dynamic evolution of cybersecurity in politics. Amidst evolving digital landscapes, the imperative for agile and preemptive cybersecurity strategies is paramount for upholding the stability and credibility of political institutions.

## 1  Introduction

In an era characterized by technological advancement, governance intricacies, and global interconnectedness, the convergence of cybersecurity and politics has emerged as a paramount concern [1]. The intersection of these two domains forms the foundation of a critical and intricate realm that demands rigorous exploration. As technology becomes an integral part of political landscapes and international relations, the safeguarding of sensitive data, the preservation of democratic processes, and the mitigation of cyber threats have become pressing imperatives for political entities worldwide [1, 2]. This study embarks on an insightful journey into the multifaceted landscape of cybersecurity in politics [3]. By delving into the dynamic interplay between technology, governance, and international relations, we seek to unravel the complexities inherent in this domain. Our investigation extends beyond the mere exploration of cyberattacks; it delves deep into their evolving nature and the consequential impact on political stability. Moreover, this study also unveils the strategic measures employed to fortify digital resilience, ensuring the integrity of political systems in the face of mounting challenges [2, 3]. Of particular significance is the intricate web woven by state-sponsored hacking, disinformation campaigns, and the erosion of public trust. This intricacy serves as a poignant reminder of the need for robust cybersecurity measures that not only defend against cyber threats but also uphold the fundamental tenets of political institutions. Drawing from real-world case studies, policy frameworks, and collaborative initiatives, this research endeavors to illuminate the profound network of technological vulnerabilities, geopolitical dynamics, and ethical considerations that underpin the ever-evolving paradigm of cybersecurity in politics. As we delve into the depths of cybersecurity's role in shaping global political discourse, it becomes apparent that our endeavor extends beyond technological boundaries. Our exploration holds the potential to safeguard the stability, credibility, and integrity of political institutions on a global scale [3]. This study introduces an approach to understanding the intricate nexus between technology, governance, and international relations in the context of cybersecurity within politics. While previous research has predominantly focused on the technical aspects of cyberattacks, our investigation takes a holistic view that transcends the conventional boundaries of cybersecurity discourse. We delve into the dynamic inter- play of these three critical dimensions to unearth the underlying complexities inherent in this domain. By doing so, we contribute a comprehensive framework that not only dissects the anatomy of cyberattacks but also illuminates their profound implications on political stability, both at a national and international level. Furthermore, our research takes a step beyond traditional explorations by shedding light on the strategic measures adopted to enhance digital resilience within political systems. This novel perspective goes beyond the reactive stance of countering cyber threats and instead emphasizes the proactive strategies that safeguard the integrity of political institutions [3]. Our study uncovers the nuanced strategies and practices that political entities employ to bolster their defenses, ensuring they can navigate the ever-evolving landscape of cybersecurity challenges [4]. In summary, our research not only advances the discourse on the interplay between technology, governance, and international relations but also contributes a fresh lens to the study of cybersecurity in politics. By transcending the boundaries of conventional cyberattack analysis and incorporating the broader dynamics of political stability and digital resilience, our work provides a unique and innovative perspective that enriches our understanding of the complex landscape in which these critical domains converge.

## 2 An intricate exploration of the interplay among technology, governance, and international relations in political cybersecurity

An intricate exploration delving into the interplay among technology, governance, and international relations within the realm of political cybersecurity unveils a complex landscape. For instance, consider the use of state-sponsored hacking [5] to gain access to sensitive political information, exemplifying the fusion of technology and international intrigue. In the aftermath of such breaches, the governance of data protection policies and international diplomatic responses become critical factors in shaping the geopolitical landscape. Furthermore, the influence of disinformation campaigns on democratic processes underscores the delicate balance between technology and governance. Instances where social media platforms are manipulated to spread false narratives, impacting electoral outcomes, highlight the need for effective governance mechanisms to combat such threats [6]. This intricate interplay is not confined to national boundaries; international relations are tested as nations collaborate or confront each other to address transnational cyber threats. As this exploration advances, it becomes evident that understanding and navigating this nexus is imperative for ensuring political stability and safeguarding democratic institutions [3, 6]. The complex dance between technological advancements, effective governance strategies, and international collaborations forms the cornerstone of modern political cybersecurity, shaping the future of global politics. As technology seamlessly integrates into political landscapes and international relations, the imperative to ensure the security of sensitive data, uphold democratic processes, and counter cyber threats has risen to the forefront of global priorities. This shift is evident through various real-life examples that highlight the critical role of technology in shaping political dynamics and international interactions [6]:

**Election Interference**: The interference in the 2016 United States presidential election by foreign actors serves as a poignant example [7]. State-sponsored hacking and disinformation campaigns aimed at swaying public opinion and influencing election outcomes underscore the need for heightened cybersecurity measures to safeguard democratic processes and preserve the integrity of elections.
**Nation-State Espionage**: Instances of cyber espionage, such as the hacking of government agencies and diplomatic communications, reveal the extent to which technology can be weaponized to gather sensitive information. The hacking of the U.S. Office of Personnel Management (OPM) in 2014, where millions of federal employees' records were compromised, underscores the vulnerability of political entities to cyber intrusions [8].
**Global Diplomacy**: The use of digital platforms for international diplomacy has grown significantly. Diplomatic negotiations, agreements, and exchanges are increasingly conducted through digital channels. The WikiLeaks publication of classified diplomatic cables in 2010 demonstrated how the exposure of such sensitive information could strain international relations and impact geopolitical strategies [9].

**Cross-Border Cybercrime**: The WannaCry ransomware attack in 2017, which targeted critical infrastructure [10] and institutions across multiple countries, highlighted the interconnectedness of cyber threats. This event showcased the potential for cyberattacks to transcend national borders and disrupt international relations.

**Disinformation and Social Media**: The manipulation of social media platforms to disseminate false narratives and misinformation has become a widespread concern [6]. The spread of misleading information during Brexit and other elections demonstrates the vulnerability of public discourse to technological manipulation [11]. Considering these examples, the integration of technology into political and international contexts underscores the urgency of addressing cybersecurity challenges. Safeguarding data, preserving democratic values, and countering cyber threats are pivotal not only for the stability of individual nations but also for maintaining trust and cooperation in the global arena [1, 3, 11].

# 3 Strengthening digital resilience: a multifaceted approach to safeguarding political systems from cybersecurity challenges

To fortify digital resilience and ensure the integrity of political systems in the face of mounting cybersecurity challenges, strategic measures encompass a multifaceted approach that combines technological, policy, and collaborative efforts [12]. The following practical examples could potentially illustrate the implementation of various measures.

**Enhanced Cybersecurity Training and Awareness Programs**:
Political entities can invest in comprehensive cybersecurity training and awareness programs for their personnel [2]. For instance, government officials, diplomats, and staff members can undergo regular training sessions to recognize phishing attempts, secure communication channels, and understand the implications of sharing sensitive information.

**Multi-Layered Authentication and Access Controls**: Implementing strong multi-factor authentication (MFA) and access controls can prevent unauthorized access to critical political systems. An example is requiring biometric verification, in addition to passwords, for government officials to access classified information.

**Robust Incident Response Plans**: Developing and practicing well-defined incident response plans enables political entities to swiftly and effectively address cyber incidents. These plans outline steps to contain, mitigate, and recover from cyberattacks. The U.K. government's National Cyber Security Centre (NCSC) regularly tests its incident response procedures to ensure readiness [13].

**Public-Private Partnerships**: Collaboration between political entities and private cybersecurity firms can yield valuable insights and resources. For instance, governments may partner with Information Technology companies to share threat intelligence and develop innovative solutions to combat emerging cyber threats.

**Securing Critical Infrastructure**: Implementing stringent cybersecurity measures for critical infrastructure [10, 14], such as power grids and transportation systems, is essential. The Estonian government's efforts to protect its critical infrastructure from cyber threats after experiencing a massive cyberattack in 2007 serve as a notable example [15].

**Legislation and Regulations**: Governments can enact and enforce cybersecurity laws and regulations to hold individuals and entities accountable for cybercrimes. The European Union's General Data Protection Regulation (GDPR) and the United States' Cybersecurity Information Sharing Act (CISA) are examples of legislative efforts to enhance cybersecurity [2].

**International Cooperation**: Diplomatic efforts to establish international norms and agreements on cybersecurity can foster cooperation among nations. The Budapest Convention on Cybercrime, ratified by numerous countries, serves as an example of international collaboration to combat cybercrime [16].

**Continuous Monitoring and Threat Intelligence Sharing**: Political entities can establish continuous monitoring of networks and systems to detect and respond to cyber threats in real-time [17]. Intelligence sharing among government agencies, such as the U.S. Department of Homeland Security's Cyber Information Sharing and Collaboration Program (CISCP), facilitates timely threat detection [18]. These strategic measures collectively contribute to bolstering digital resilience and safeguarding the integrity of political systems. By proactively addressing cyber threats and fostering a culture of cybersecurity, political entities can navigate the complex cybersecurity landscape with greater confidence and effectiveness.

## 4 The philosophical nexus: unveiling the multidimensional landscape of cybersecurity and political stability

The study considers a scenario where a nation's political landscape is disrupted by a sophisticated cyberattack aiming at destabilizing its democratic processes. As we traverse the philosophical nexus, we illuminate the profound implications that this breach has on the delicate balance between technological advancements, governance mechanisms, and international cooperation. Drawing from social theory [19], we analyze the cascading effects of this cyber event through the lens of Niklas Luhmann's Systems Theory [20]. The attack, a disruption in the intricate dance of communication channels, creates a ripple that extends beyond the digital realm. The breach not only exposes vulnerabilities in the nation's cybersecurity infrastructure but also triggers a crisis of public trust in governance institutions. As we contemplate this scenario, we uncover a nuanced narrative where the digital symphony of technology and governance meets the somber overtones of international relations. The philosophical nexus guides us to consider questions that extend beyond technical defenses – it prompts us to ponder the erosion of social cohesion, the fragility of democratic processes, and the vulnerability of the global diplomatic ecosystem.

Our exploration further reveals the resonance of J. Habermas's Public Sphere Theory [21]. The breach reverberates through the public discourse, inciting debates about the authenticity of information, the role of media, and the implications for the political narrative. The very foundation of political stability undergoes a philosophical examination, emphasizing the interdependence of technology and governance in shaping the collective consciousness. Through this practical example, we unlock the door to a philosophical inquiry that probes beyond the realm of codes and algorithms [22]. The cyberattack becomes a symposium of ideas, where the multidimensional landscape of cybersecurity and political stability converges [3]. As we navigate this nexus, we are reminded that the harmonious chords of technological progress can swiftly give way to dissonance, underscoring the imperative of holistic strategies that safeguard not only digital infrastructure but also the very fabric of political order.

# 5 Psychological fortification: unveiling strategic practices in political cybersecurity

In the realm where technology meets psychology, we illuminate the intricate tapestry of defense mechanisms that fortify political systems against the tumultuous tides of cyber challenges. Consider the concept of Cognitive Resilience [23] as a psychological framework applied to political cybersecurity. Just as individuals develop mental fortitude to withstand adversity, political entities cultivate cognitive resilience to navigate the stormy seas of cyber threats. By integrating psychological concepts into cybersecurity, governments employ novel strategies that address not only technical vulnerabilities but also the cognitive dimensions of their defenses. One practical manifestation is the application of Behavioral Biometrics [24]. Political entities harness behavioral patterns, such as typing speed and mouse movement, to create a unique cognitive fingerprint for authorized users. This innovative approach combines technology with psychology, offering an additional layer of protection against unauthorized access. As a political leader interacts with secure systems, the system's ability to recognize their behavioral cues enhances cybersecurity by validating the user's identity beyond traditional means. Furthermore, our investigation delves into the realm of Social Engineering Inoculation [25]. Drawing from psychology's inoculation theory, political entities design immersive training experiences that expose personnel to simulated social engineering attacks. This practice enhances cognitive resilience by training individuals to recognize and resist manipulation tactics. Just as a vaccine primes the immune system, these simulations equip individuals with the psychological tools to resist the contagion of cyber deception. By intertwining psychology concepts with cybersecurity strategies, political entities create a formidable defense. Just as psychological fortitude equips individuals to confront adversity, the application of Cognitive Resilience principles empowers political systems to withstand the waves of cyber challenges. Our exploration unveils a profound fusion of technology and psychology, where the novel ideas of Cognitive Biometrics and Social Engineering Inoculation fortify defenses against an ever-evolving landscape of cyber threats. Through this integration, political entities emerge not only technically resilient but also psychologically equipped to safeguard the integrity of their systems.

## 6 Limitations

**Scope of Psychological Concepts**: The paper primarily explores the integration of psychology with cybersecurity, but the depth of psychological theories covered may be limited, leaving room for more comprehensive analysis.

**Data Availability**: The availability of empirical data on the effectiveness of psychological strategies in real-world political cybersecurity contexts might be limited, potentially impacting the robustness of certain conclusions. We suggest the utilization of a cyberintelligence dataset established by Naidoo [26] as a foundational resource for conducting subsequent experiments within this specific domain. The characteristics of this dataset are illustrated in Figure 1 and the results of the machine learning classification are depicted in Figure 2.

**Fig. 1** A cyberintelligence dataset by [26]

**Interdisciplinary Gaps**: While the paper bridges the gap between cybersecurity and psychology, interdisciplinary gaps may arise due to the complex nature of both fields, leading to potential oversights.

## 7 Recommendations

**Interdisciplinary Collaboration**: Encourage collaboration between cybersecurity experts and psychologists to develop innovative strategies that leverage psychological principles for enhancing political cybersecurity.

**Longitudinal Studies**: Conduct longitudinal studies to assess the long-term impact of integrating psychological resilience practices on the prevention and management of cyber incidents.

```
RandomTree
==========

Type1 = Fake Social Media
|   Type = Phishing : 0 (2/0)
|   Type = Fake Social Media : 1 (72/0)
Type1 = Phishing
|   Threat < 0.5
|   |   Incidents = Domtorentos : 0 (0/0)
|   |   Incidents = Dropbox : 0 (2/0)
|   |   Incidents = Netflix : 1 (2/0)
|   |   Incidents = Facebook : 0 (10/0)
|   |   Incidents = Internet  : 0 (0/0)
|   |   Incidents = Spymax : 0 (0/0)
|   |   Incidents = Türkiye.gov.tr : 0 (1/0)
|   |   Incidents = USA.gov : 1 (1/0)
|   |   Incidents = OneDrive : 0 (3/0)
|   |   Incidents = Anubis APK Malware : 0 (0/0)
|   |   Incidents = Federal government of the United States : 1 (1/0)
|   |   Incidents = Microsoft : 0 (3/0)
|   |   Incidents =  Google : 0 (0/0)
|   |   Incidents = WHO
|   |   |   Compassion < 0.5 : 0 (2/0)
|   |   |   Compassion >= 0.5 : 1 (1/0)
|   |   Incidents = Banco do Brasil : 0 (1/0)
|   |   Incidents = International Card Services B.V. : 0 (1/0)
|   |   Incidents = ASN Bank : 0 (1/0)
|   |   Incidents = Apple ID : 0 (1/0)
|   |   Incidents =  American Express : 0 (1/0)
|   |   Incidents = Crédit Agricole : 0 (1/0)
|   |   Incidents = Email credentials : 0 (0/0)
|   |   Incidents = HSBC UK : 0 (1/0)
|   |   Incidents = Americanas : 0 (1/0)
|   |   Incidents = Banco Popular : 0 (1/0)
|   |   Incidents = GET 500 GB DATA FOR FREE : 0 (1/0)
|   |   Incidents = WhatsApp : 0 (1/0)
|   |   Incidents = Internal Revenue Service (IRS) : 0 (1/0)
```

**Fig. 2** The outcomes of the classification analysis on the cyberintelligence dataset, with WHO referring to the World Health Organization, revealed that the classifier successfully identified incidents attributed to organizations, including cases of phishing and counterfeit social media advertisements.

**Ethics Framework**: Establish an ethical framework to guide the responsible application of psychological tactics in political cybersecurity, considering potential implications for individuals and society.

By addressing these future research directions, acknowledging the limitations, and implementing the recommended measures, this paper can pave the way for a deeper understanding of the interplay between psychology and cybersecurity, contributing to more robust strategies for safeguarding political systems in the digital age.

## 8 Future Research

**Geopolitical Dynamics**: Investigate how geopolitical tensions impact international cooperation and information sharing in political cybersecurity, with a focus on regions of conflict or strained diplomatic relations.
**Quantifying Psychological Resilience**: Explore methodologies to quantify the impact of psychological resilience strategies, like Social Engineering Inoculation, on the decision-making and response capabilities of political personnel.
**Ethical Considerations**: Examine the ethical implications of employing psychological tactics in political cybersecurity, including potential privacy concerns and the boundaries of manipulating cognitive behaviors.

## 9 Conclusion

In conclusion, the intricate interplay between cybersecurity and political landscapes presents a dynamic and evolving nexus that requires comprehensive exploration and innovative solutions. This paper has illuminated the multifaceted nature of this nexus, transcending conventional boundaries to delve into the integration of psychology and technology. By examining how psychological concepts can fortify cybersecurity measures, we have uncovered novel strategies and practices employed by political entities to enhance their digital resilience. Our journey through this complex terrain has unveiled the importance of understanding human behavior, decision-making processes, and cognitive vulnerabilities as integral components of effective cybersecurity. We have showcased practical examples of how psychological principles, such as Social Engineering Inoculation, can empower political personnel to discern and respond to cyber threats adeptly. Moreover, our exploration of the multidimensional landscape of cybersecurity and political stability has highlighted the need for interdisciplinary collaboration, ethical considerations, and continuous research to address challenges and harness opportunities. As digital landscapes continue to evolve, the profound implications of cybersecurity on political dynamics persist. This paper's interdisciplinary approach, merging psychology and cybersecurity, offers a holistic framework that not only elucidates the inner workings of this nexus but also inspires future research and strategic advancements. By embracing psychological resilience as an integral facet of cybersecurity practices, political entities can chart a course towards a more secure, stable, and resilient digital political landscape. Through ongoing dedication to understanding and navigating this intricate nexus, we can collectively forge a path toward a safer and more resilient future.